# Enhanced photocurrent efficiency of a carbon nanotube electromagnetically coupled to a photonic structure

*Bryan M. Wong and Alfredo M. Morales*

Materials Chemistry Department, Sandia National Laboratories, Livermore, California 94551, USA

E-mail: bmwong@sandia.gov

**Abstract.** We present photocurrent power-enhancement calculations of a carbon nanotube *p-n* junction electromagnetically coupled to a highly-efficient photonic structure. Particular attention is paid to a GaAs photonic structure specifically modified to increase the intensity of infrared light onto the nanotube region for effective energy conversion. Using finite-difference time-domain calculations, we compute a significant increase in electric field intensity in the nanotube region which enables an estimation of power efficiency. These results demonstrate the potential of using a photonic structure to couple large-scale infrared sources with carbon nanotubes while still retaining all the unique optoelectronic properties found at the nanoscale.

**Introduction**

One of the most rapidly-growing areas in nanoscience is the ability to artificially manipulate optical and electrical properties at the nanoscale. In particular, nanomaterials such as single-wall carbon nanotubes can convert infrared light into photocurrent due to their unique one-dimensional electronic properties [1]. Indeed, recent experimental measurements have shown that a single carbon nanotube can act as a polarized photodetector with quantum efficiencies greater than 10% [2]. In addition to intrinsic optoelectronic efficiency, carbon nanotubes exhibit ballistic charge transport at room temperature [3], potentially enabling high-performance photodetectors that need not be cryogenically cooled.

While numerous theoretical and experimental studies have been carried out to exploit the electronic properties of individual carbon nanotubes, it has been difficult to efficiently couple microscale technologies to the nanoscale properties of individual nanotubes. In particular, one of the most limiting factors in making nanotube-based photodetectors a viable technology is their relatively small absorption cross-section. Since the diameter of a nanotube is on the order of 1 nm, its light-collecting surface area is limited to this small length scale, and radiation which is not directly incident on the nanotube will go undetected [4,5].

To address this problem, we investigate the photocurrent efficiency of a carbon nanotube when it is illuminated by a highly-efficient photonic structure. This coupling of carbon nanotubes with photonic structures is a natural and logical step towards bridging microscale and nanoscale length scales. By artificially modifying the composition and periodic arrangement in a photonic structure, the flow of light can be guided onto extremely localized regions of space. Recognizing these possibilities, several groups have already characterized photonic structures which can focus electromagnetic waves and provide highly-directional light beams [6-14]. There has also been recent work in modifying photonic surfaces to produce beams with high transmission efficiencies [8]. In this work, we specifically calculate the efficiency of a carbon nanotube *p-n* junction which is electromagnetically coupled to a planar GaAs photonic structure. To validate the performance of the photonic system, finite-difference time-domain (FDTD) calculations are used to calculate electric field enhancements, power intensities, and coupling efficiencies. The resulting FDTD calculations enable an estimation of nanotube photocurrent efficiency within a linear response approximation and also provide guidance on enhancing this efficiency using a photonic structure.

**Photonic Structure Specifications**

The two-dimensional photonic structure considered in this work is based on the design of Lin et al. [12] who originally used a silicon structure to couple terahertz radiation into a narrow waveguide. For the nanotube-photonic structure in this work, we are primarily interested in the infrared region of the electromagnetic spectrum, so a few modifications have been made to their original design. The most significant and important modification is the use of a GaAs ($\varepsilon = 13.2$) photonic substrate which is largely transparent to infrared light [15]. The photonic structure studied here is a periodic square array of air holes arranged on a 48 $\mu$m × 85 $\mu$m GaAs slab shown in figure 1. The entire photonic structure is enclosed with absorbing boundary layers with 15 $\mu$m of air ($\varepsilon = 1.0$) between the top boundary layer and the cleaved GaAs slab. The ratio of the air hole radius to the lattice spacing is $r/a = 0.3$ where $a = 1.365$ $\mu$m. The corresponding band structure of this periodic array is shown in figure 2 for all wave vectors along the edges of the irreducible Brillouin zone defined by the points Γ, M, and X. Using a plane-wave expansion method [16], the resulting band structure was obtained for a transverse magnetic (TM) mode which has its electric field perpendicular to the plane of the GaAs slab. An effective index of 3.0 was used for the background GaAs material which corresponds to a guided mode in the slab having a wavelength of around 10.6 $\mu$m. From the band structure, it can be deduced that a wavelength of 10.6 $\mu$m lies in a frequency range contained in the first band for incident light along any direction. By choosing the incident light to coincide with a frequency in the first band, any plane wave launched into the photonic structure will propagate with minimal transmission loss. Once the light has been guided through the photonic surface, the propagation path of the incident plane wave will be bent and altered by the pattern of air holes in the GaAs substrate.

To investigate the focusing effects of the photonic structure, a two-dimensional FDTD method was used to propagate Maxwell's equations in time and space throughout the entire structure [13]. A 50 $\mu$m-wide input wave (10.6 $\mu$m wavelength) was launched along the ΓX direction into the photonic structure, and the steady-state electric field intensities ($\propto |E_z|^2$) are shown in figure 3. After passing through the array of air holes, the light propagates over several microns, subsequently focuses, and finally exits the GaAs material into air. The performance of the photonic structure can be quantitatively characterized by calculating the $E_z$ intensity along a vertical line through the center of the structure. As shown in figure 3, the photonic structure tightly focuses the wide plane wave into a narrow 5-$\mu$m-wide region which lies 0.1 $\mu$m outside the cleaved GaAs facet. Along this line scan, the electric field intensity reaches a maximum value of 6.25 relative to the source intensity which was set to a value of 1.0. From the $E_z$ and $B_x$ field distributions, a coupling efficiency of 63% was obtained by integrating the $y$-component of the Poynting vector across a 10 $\mu$m wide region near the focusing point and dividing it by a similar integration across the entire 50 $\mu$m source region. In other words, 63% of the input power due to the 50 $\mu$m-wide light wave is focused in a 10 $\mu$m wide region near the GaAs-air interface.

It is important to mention that we also performed FDTD calculations on an uncleaved GaAs slab without the 15 $\mu$m air interface and found that the intensity was significantly reduced by a factor of 2.5 compared to the present photonic structure. The increased intensity of the present configuration is mostly due to the rapid change in permittivity as the light waves approach the GaAs-air interface. Once the light finally exits the GaAs material into air, it slightly refocuses near the interface again before spreading out into free space.

**Carbon Nanotube Photocurrent Calculations**

The final step in evaluating the efficiency of this nanotube-photonic device is to calculate the increase in nanotube photocurrent due to the photonic structure. As mentioned previously, a carbon nanotube acts as a polarized photodetector and generates maximal photocurrent when the incident radiation has an electric field which is parallel to the nanotube axis. Since the photonic structure is operating in the TM mode with an $E_z$ component perpendicular to the plane of the GaAs slab, the carbon nanotube was placed near the GaAs-air interface with its axis parallel to the $E_z$ field for maximum efficiency.

The theoretical performance of a nanotube *p-n* junction under bias has been described by Stewart and Leonard [17]. Their calculation is based on a self-consistent nonequilibrium Green's function approach to determine current-voltage characteristics of illuminated carbon nanotubes under an applied voltage bias. Under low power conditions, one can use Stewart and Leonard's expressions for the efficiency $\eta$ of a nanotube *p-n* junction under bias given by [17]

$$\eta = \frac{kT}{4eI_s} \frac{I_{ph}^2}{P_{op}} \qquad (1)$$

where $I_s$ is the saturation current of the nanotube, $I_{ph}$ is the current generated by the electron-photon interaction, and $P_{op}$ is the optical power delivered to the nanotube. Since $I_{ph}$ depends linearly on $P_{op}$, equation (1) reduces to $\eta \propto P_{op}$ in the low optical power limit. Therefore, a relative intensity of 6.25 obtained from the FDTD calculations directly translates into a factor of 6.25× enhancement (or alternatively, a 525% increase) in photocurrent efficiency. While we have specifically focused on a GaAs photonic slab structure, it is clear that further study on merging these two technologies need further study. To our knowledge, the experimental photocurrent of a carbon nanotube electromagnetically coupled to a photonic structure has not been studied and remains unknown. Experimental measurements on the photocurrent power efficiency combining these two technologies would be extremely valuable as a check on our theoretical calculations.

**Conclusion**

In conclusion, we have investigated the feasibility of using a photonic structure to enhance the photocurrent efficiency of a carbon nanotube device. Using a two-dimensional FDTD method, we designed a GaAs photonic structure which modulates the propagation of infrared light to focus plane waves into a 5 $\mu$m-wide region. The subsequent increase in electric field intensity in this region enables the computation of power efficiency within a linear response approximation.

Using a low-power approximation for the carbon nanotube junction response, the photonic structure thus designed enables a 525% increase in photocurrent. Photonic structures may also be combined with other chemical methods which direct and control the electronic properties of the nanotubes themselves to further enhance the photocurrent efficiency. One promising method already under study is the noncovalent chemical functionalization of carbon nanotubes [18].

**Acknowledgement**

Supported by the Laboratory Directed Research and Development program at Sandia National Laboratory, a multiprogram laboratory operated by Sandia Corporation, a Lockheed Martin Company, for the United States Department of Energy's National Nuclear Security Administration under contract DE-AC04-94AL85000.

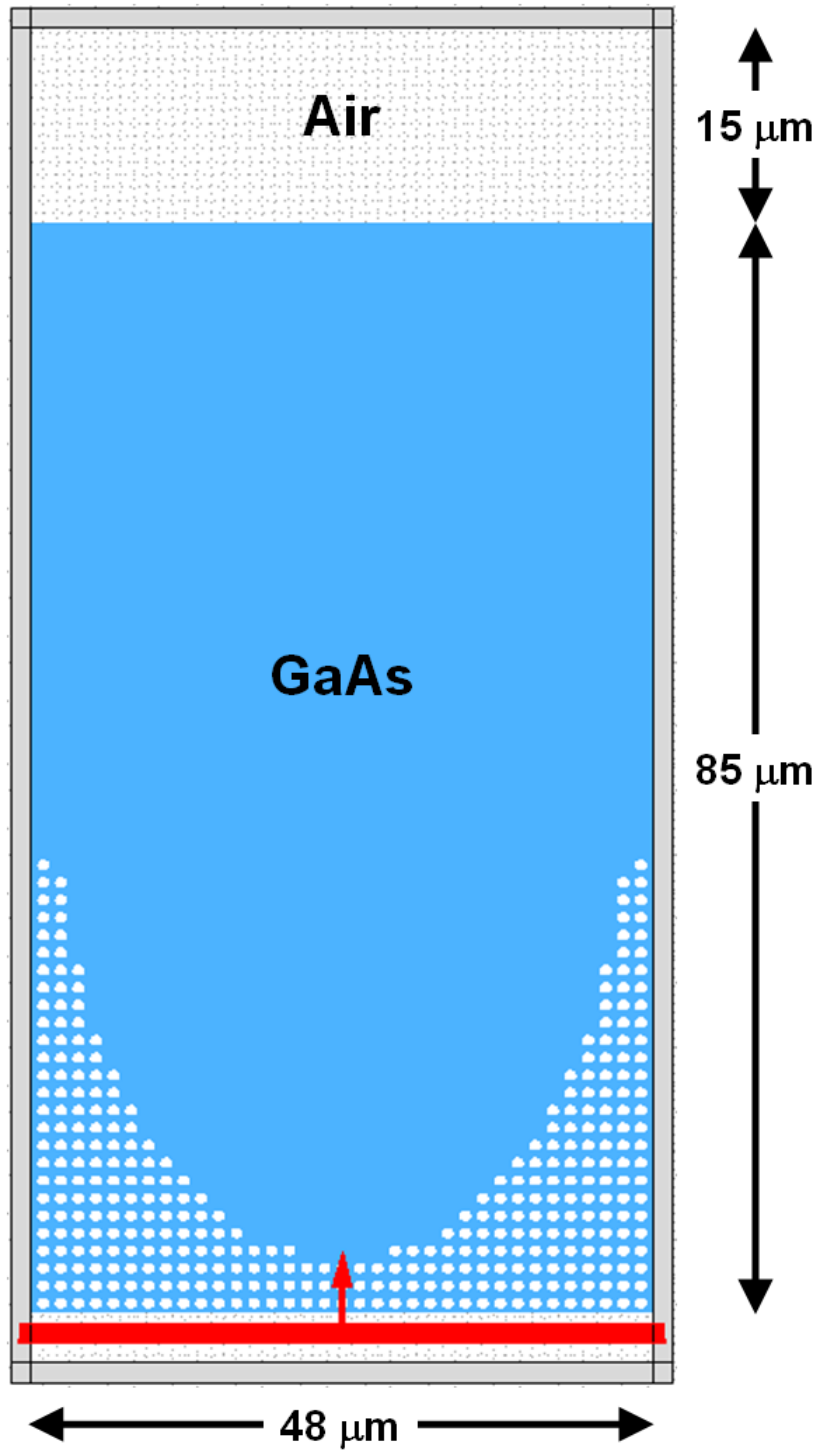

**Figure 1.** GaAs photonic structure enclosed with absorbing boundary layers. The ratio of the air hole radius to the lattice spacing is $r/a = 0.3$ where $a = 1.365$ μm. A plane wave approaching the structure from below is shown in red.

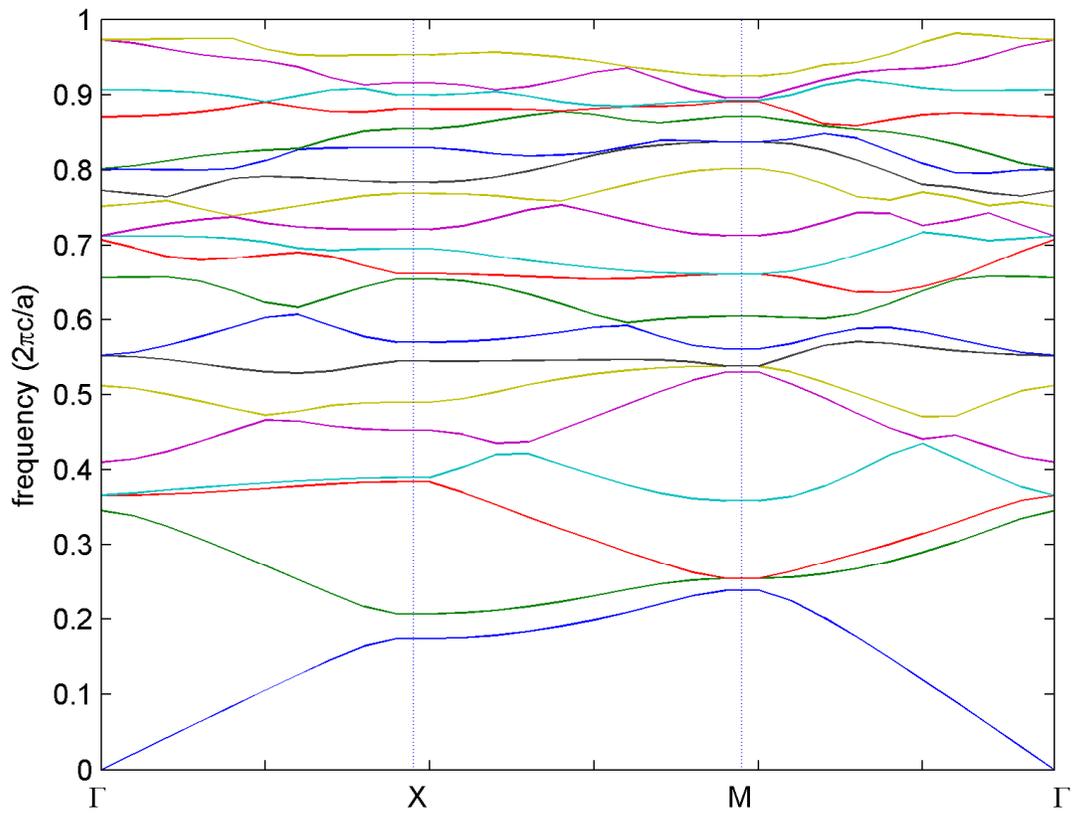

**Figure 2.** Calculated photonic band structure of the GaAs photonic array. A wavelength of 10.6 $\mu$m lies in a frequency range contained in the first band for incident light along any direction.

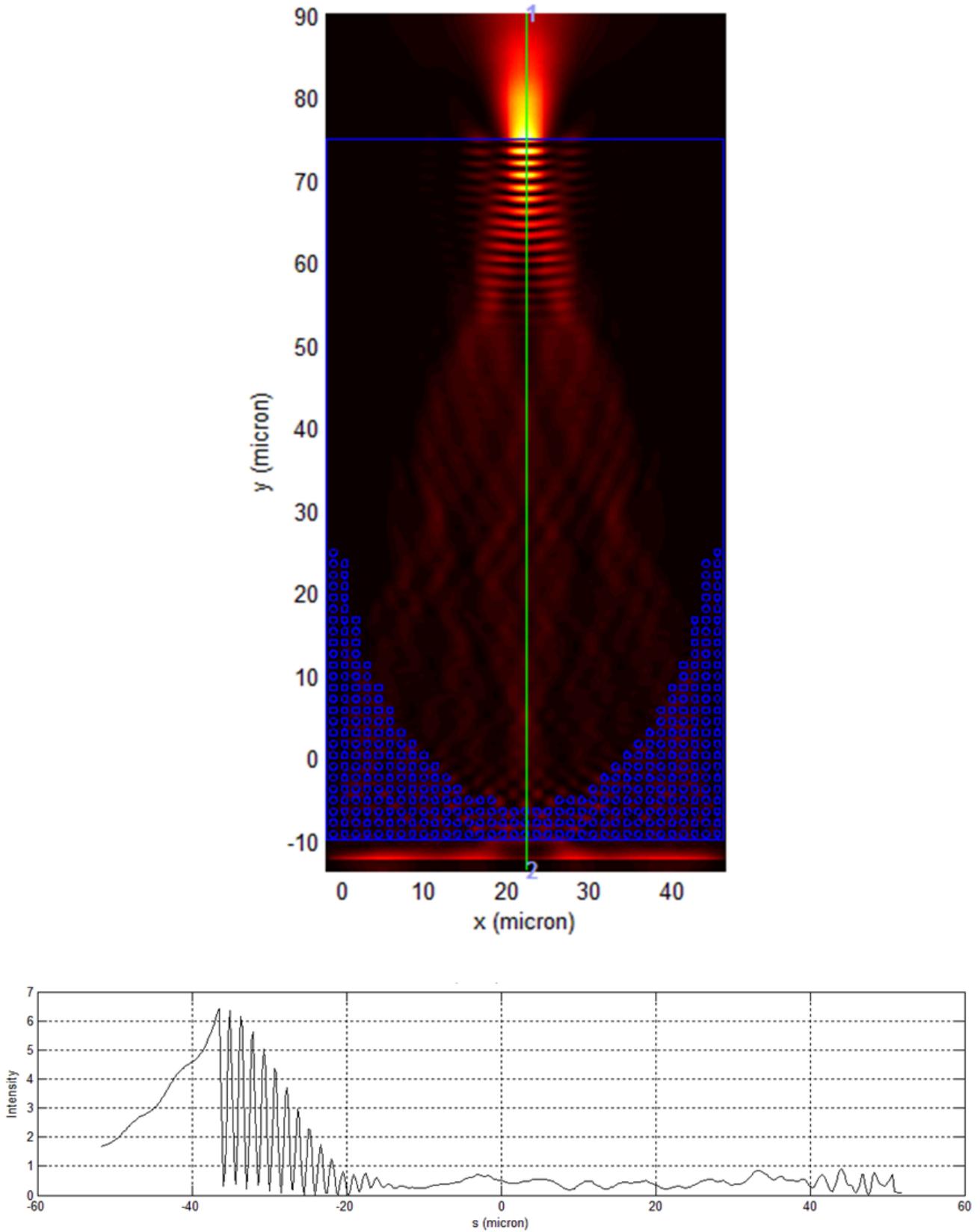

**Figure 3.** Intensity profile of the GaAs photonic structure obtained from a two-dimensional FDTD analysis. The z-component of the electric field intensity along the vertical green line is shown below. The source intensity is normalized to a value of 1.0.

[1] Saito R, Dresselhaus G and Dresselhaus M S 1998 *Physical Properties of Carbon Nanotubes* (London, UK: Imperial College Press)

[2] Freitag M, Martin Y, Misewich J A, Martel R and Avouris Ph 2003 Nano Lett. **3** 1067

[3] Javey A, Guo J, Wang Q, Lundstrom M and Dai H 2003 Nature **424** 654

[4] Burke P J, Li S and Yu Z 2006 IEEE Trans. Nanotechnol. **5** 314

[5] Kempa K, Rybczynski J, Huang Z, Gregorczyk K, Vidan A, Kimball B, Carlson J, Benham G, Wang Y, Herczynski A and Ren Z 2007 Adv. Mater. **19** 421

[6] Ozbay E, Aydin K, Bulu I and Guven K 2007 J. Phys. D: Appl. Phys. **40** 2652

[7] Li Z, Aydin K and Ozbay E 2008 J. Phys. D: Appl. Phys. **41** 155115

[8] Wang Q, Cui Y, Yan C, Zhang L and Zhang J 2008 J. Phys. D: Appl. Phys. **41** 105110

[9] Vodo P, Parimi P V, Lu W T and Sridhar S 2005 Appl. Phys. Lett. **86** 201108

[10] Qiu M, Thylén L, Swillo M and Jaskorzynska B 2003 IEEE J. Quantum Electron. **9** 106

[11] Berrier A, Mulot M, Swillo M, Qiu M, Thylén L, Talneau A and Anand S 2004 Phys. Rev. Lett. **93** 073902-1

[12] Lin C, Chen C, Sharkawy A, Schneider G J, Venkataraman S and Prather D W 2005 Opt. Lett. **30** 1330

[13] Sharkawy A, Pustai D, Shi S, Prather D W, McBride S and Zanzucchi P 2005 Opt. Express. **13** 2814

[14] Prather D W, Shi S, Murakowski J, Schneider G J, Sharkawy A, Chen C and Miao B 2006 IEEE J. Quantum Electron. **12** 1416

[15] Brozel M R and Stillman G E 1996 *Properties of Gallium Arsenide*, 3rd ed. (London, UK: The Institution of Electrical Engineers)

[16] Joannopoulos J, Mead R and Winn J, 1995 *Photonic Crystals* (Princeton, NJ: Princeton Univ. Press)

[17] Stewart D A and Léonard F 2005 Nano Lett. **5**, 219

[18] Zhou X, Zifer T, Wong B M, Krafcik K L, Léonard F and Vance A L "Color Detection Using Chromophore-Nanotube Hybrid Devices" Nano Lett. (in press)